\documentclass[12pt]{article}
\usepackage[english]{babel}
\usepackage{amsmath}
\usepackage{amssymb}
\usepackage{bbm}
\usepackage{color}
\usepackage[titletoc,toc,title]{appendix}
\Roman{section}
\newcommand{\eq}[1]{\begin{equation}#1\end{equation}}
\newcommand{\naw}[1]{\left(#1\right)}
\newcommand{\ket}[1]{\left|#1\right>}
\newcommand{\bra}[1]{\left<#1\right|}
\newcommand{\av}[1]{\left<#1\right>}
\newcommand{\com}[1]{\left[#1\right]}
\newcommand{\modu}[1]{\left|#1\right|}

\begin{document}

\begin{center}
\textsc{\Large{General quantum two-players games, their gate operators and Nash equilibria}}

\emph{Katarzyna Bolonek-Laso\'n\footnote{kbolonek1@wp.pl}\\ Department of Statistical Methods, Faculty of Economics and Sociology, \\ University of Lodz\\
41/43 Rewolucji 1905 St., 90-214 Lodz, Poland.}

\end{center}
\begin{abstract}
The two-players $N$ strategies games quantized according to the Eisert-Lewenstein-Wilkens scheme (\emph{Phys. Rev. Lett.} \textbf{83} (1999), 3077) are considered. Group theoretical methods are applied to the problem of finding a general form of gate operators (entanglers) under the assumption that the set of classical pure strategies is contained in the set of pure quantum ones.

The role of the stability group of the initial state of the game is stressed. As an example, it is shown that the maximally entangled games do not admit nontrivial pure Nash strategies. The general arguments are supported by explicit computations performed in the three strategies case. 
\end{abstract}

\section{Introduction}
In two important papers \cite{EisertWL}, \cite{EisertW} Eisert, Wilkens and Lewenstein proposed the method which allows, given some classical non-cooperative game, to construct its quantum counterpart. The example they described provides a paradigm of a wide class of quantum games. Since then the theory of quantum games has been a subject of intensive research  \cite{BenjaminHay}$\div$\cite{Brunner}.

In their attempt to justify the interest in quantum games Eisert, Lewenstein and Wilkens speculate that games of survival are being played already on molecular level where things are happening according to the rules of quantum mechanics. They also pointed out that there is an intimate connection between the theory of games and the theory of quantum  communication.

The Eisert-Lewenstein-Wilkens (ELW) game can be played by purely classical means. To this end one can compute (on classical computer), according to the standard rules of quantum theory, the relevant probabilities (and payoffs) and toss coins which are appropriately biased on these values. However, it can happen that this is not physically feasible due to limited resources and time. In such a case only quantum mechanics allows for an implementation of the game due to the existence of specific quantum correlations which, in general, break the Bell-like inequalities. In this respect quantum games resemble quantum coding or quantum computing: the use of non-classical correlations can lead to high effectiveness.

Let us briefly describe the original ELW proposal \cite{EisertWL}. One starts with classical two-players (Alice and Bob) two-strategies ($C$ (cooperate) and $D$ (defect)) non-cooperative symmetric game described in  Table 1.
\begin{table}
\caption{The payoffs resulting from different EWL strategies.}
\begin{center}
\begin{tabular}{|c|c|c|c|}\hline
\multicolumn{2}{|c|}{Strategies} & \multicolumn{2}{|c|}{Payoffs}\\
\cline{1-4}
player A & player B  & player A & player B\\
\cline{1-4}
C & C & r & r\\
C & D & s & t\\
D & C & t & s\\
D & D & p & p\\ \hline
\end{tabular}
\end{center}
\end{table}

 The quantization of the classical game described by the above table begins by assigning the possible outcomes of the classical strategies C  and D to the basis vectors $\ket{1}$ and $\ket{2}$ of twodimensional complex Hilbert space.  The state of the game is described by a vector in the tensor product space spanned by the vectors $\ket{1}\otimes\ket{1}$, $\ket{1}\otimes\ket{2}$, $\ket{2}\otimes\ket{1}$ and $\ket{2}\otimes\ket{2}$ which correspond to all possible choices of both players classical strategies. The initial state of the game is given by
\eq{\ket{\Psi_{in}}=J\naw{\ket{1}\otimes\ket{1}},}
where $J$ is a unitary operator known to both players.  $J$ plays the crucial role because it introduces the entanglement allowing for the genuinely quantum correlations. It is called the gate operator or entangler.
 Strategic moves of both players are associated with unitary $2\times 2$ operators $U_A$, $U_B$ operating on their own qubits. The resulting final state of the game is given by
\eq{\ket{\Psi_{out}}=J^+\naw{U_A\otimes U_B}\ket{\Psi_{in}}=J^+\naw{U_A\otimes U_B} J\naw{\ket{1}\otimes\ket{1}}.}
Denoting
\eq{P_{kk'}\equiv\modu{\av{k\otimes k'|\Psi_{out}}}^2,\qquad k, k'=1,2\label{b}}
the expected payoffs are computed according to 
\eq{\begin{split}
& \$_A=rP_{11}+pP_{22}+tP_{21}+sP_{12}\\
&  \$_B=rP_{11}+pP_{22}+sP_{21}+tP_{12}
\end{split}.\label{abc}}

There are three main elements which determine the properties of ELW game.
\begin{description}
\item{(i)} First, one chooses the classical payoff table, i.e. the values $p$, $r$, $s$ and $t$. The classical game is then uniquely defined. Some choices are more interesting than others. For example, if the classical payoffs obey $t>r>p>s$, the Prisoner Dilemma emerges on the classical level.
\item{(ii)} The crucial role is played by the gate operator $J$ (entangler) which introduces quantum entanglement. It converts the classical game into genuinely quantum one. Two assumptions are made concerning the form of $J$: (a) to preserve the symmetry of the game $J$ is symmetric with respect to the interchange of the players; (b) the quantum game entails a faithful representation of its classical counterpart. In the case of original ELW game (a) and (b) determine $J$ up to one free parameter; namely,
\eq{J=\exp\naw{-\frac{i\gamma}{2}\sigma_2\otimes\sigma_2}\label{a},}
where $\gamma$ is real and $\sigma_2$ is the second Pauli matrix.
\item{(iii)} The properties of the ELW game depend also on the choice of the subset $\Sigma$ of allowed strategies $U_A$ and $U_B$. In general, $\Sigma\subset SU(2)$ because the trivial $U(1)$ factor can be neglected. In the original Eisert et al. proposal the allowed strategies belong to the twodimensional submanifold of $SU(2)$ which itself is not a group. This point of view was criticized by Benjamin and Hayden \cite{BenjaminHay} who pointed out that there are no compelling reasons to impose such a restriction; it seems difficult to find a physical justification for the choice proposed by Eisert et al. We shall adopt the point of view presented in Ref. \cite{BenjaminHay} and assume that the manifold of admissible strategies forms always a group.
\end{description} 

The aim of the present paper is twofold. We generalize the ELW construction to the case of two-players N-strategies games. Again, the starting point is a noncooperative classical game defined by an arbitrary symmetric payoffs table. Quantum strategies of Alice and Bob are represented by arbitrary unitary matrices (neglecting irrelevant overall phase factor), i.e. we assume $\Sigma = SU(N)$. The only nontrivial point consists in defining an appropriate entangler $J$. We demand, following original ELW construction, that the resulting guantum game is symmetric and includes the classical game. It appears then that there exists a multiparameter family of acceptable entanglers $J$ with the number of arbitrary parameters growing quadratically with $N$. As a result we obtain a far reaching generalization of the original ELW game.

 Our second aim is to show that the group theoretical methods provide quite powerful tool for analyzing the general properties of quantum games. A good example is provided by the construction of the entangler $J$ which is based on considering the cyclic subgroup of permutation group. Next we show that an important role is played by the stability group of initial state of the game. Its structure depends, to some extent, on the entanglement degree of $\ket{\Psi_{in}}$; the maximally entangled state corresponds to the  large stability group. As a result maximally entangled games have peculiar properties. To see this consider the $N=2$ case. The relevant entangler is given by eq. (\ref{a}). The case of maximal entanglement corresponds to $\gamma=\frac{\pi}{2}$. It has been shown by Landsburg \cite{Landsburg}, \cite{Landsburg1}, \cite{Landsburg2} that for this value of $\gamma$ the game can be described in terms of quaternions algebra. Moreover, the resulting outcome probabilities depend only on the product of quaternions representing the strategies of Alice and Bob. This allows us to conclude, for example, that no nontrivial (in the sense described below) pure Nash equilibrium exists. It has been shown in Ref. \cite{Bolonek} that the quaternionic structure (and the real Hilbert space structure behind it) and nonexistence of Nash equilibria result from the structure of stability group of initial vector.  In the present paper we generalize this result. Although for $N>2$ the quaternionic structure of the quantum game is lost one can still show that, in the case of maximal entanglement, no nontrivial pure Nash equilibrium exists. This result is very general. It depends neither on the form of classical payoff table nor on the actual form of gate operator. The proof is very simple and based on group theoretical considerations. It shows the power of group theory methods.

The paper is organized as follows. In Sec. II we describe the generalization of ELW game to the case of N strategies. Then we prove that no nontrivial pure Nash equilibrium exists if the initial state is maximally entangled. 

In Sec. III a wide class of entanglers is constructed for arbitrary N. The construction is based on simple use of the representation of cyclic subgroup of permutation group. It is shown that the number of free parameters is essentially determined by the rank of $SU(N)$ and is proportional to $N^2$.

The case $N=3$ is considered in more detail in Sec. IV. The general three parameter gate operator is explicitly constructed. All values of the parametes leading to maximally entangled games are determined. Some non-maximally entangled games are considered which correspond to doubly degenerated or nondegenerated initial reduced density matrices. In a number of cases the explicit form of generators of stability group is determined and shown to agree with the general results obtained in Sec. III.

Sec. V is devoted to some conlusions. A number of technical details is relegated to the Appendices.

The present work is based on three papers \cite{Bolonek1}, \cite{Bolonek2}, \cite{Bolonek3} which appeared on arXiv.

\section{The two-players N-strategies quantum games}

The original ELW construction of quantum game can be generalized as follows. The starting point is some classical noncooperative 2-players N-strategies symmetric game defined by a relevant payoff table. In order to construct its quantum version one ascribes to any player (Alice and Bob) an N-dimensional complex Hilbert space spanned by the vectors 
\begin{equation}
\ket{1}=\left(\begin{array}{c}
1 \\
0 \\
\vdots \\
0 \end{array}\right),\quad \ldots, \quad \ket{N}=\left(\begin{array}{c}
0\\
\vdots\\
0 \\
1
\end{array}\right ).\label{ab}
\end{equation}  

One starts with the vector $\ket{1}\otimes\ket{1}$. The entanglement of  initial state is provided by a reversible gate operator $J$ (entangler); therefore,
\begin{equation}
\ket{\Psi_{in}}\equiv J\naw{\ket{1}\otimes\ket{1}}\label{a1}
\end{equation}
is the initial state of the game, where now $\ket{1}$ refers to the first vector in eqs. (\ref{ab}). In the present section the explicit form of $J$ is not relevant. We only assume that $J$ is symmetric with respect to the permutation of the factors entering the tensor product (to preserve the symmetry of the game) and the classical game is faithfully represented in its quantum counterpart.

We assume further that the set of allowed strategies, both for Alice and Bob, is the whole $SU(N)$ group (the overall phase can be factored out and becomes irrelevant). The players perform their moves and then the final measurement is made yielding the final state of the game
\begin{equation}
\ket{\Psi_{out}}=J^+\naw{U_A\otimes U_B}J\naw{\ket{1}\otimes\ket{1}}.
\end{equation} 
This allows us to compute the players expected payoffs:
\begin{equation}
\$^{A,B}=\sum_{k,k'=1}^{N}p_{k,k'}^{A,B}\modu{\av{k,k'|\Psi_{out}}}^2,
\end{equation}
where $\ket{k,k'}\equiv\ket{k}\otimes\ket{k'}$, $k,k'=1,...,N$ and $p_{k,k'}^{A,B}$ are classical payoffs of Alice and Bob, respectively.

We see that the construction of generalized ELW game proceeds along the same lines as in the original $SU(2)$ case. There is, however, an important difference. Since the $SU(2)$ group has rank one, the set of allowed gate operators $J$ is parametrized by one real parameter $\gamma$ (cf. eq. (\ref{a})). For general $N$ there is much more freedom for the choice of $J$. In fact, as it will be shown below, $J$ depends on a number of free parameters growing proportionally to $N^2$. However, the explicit form of $J$ is irrelevant for the problem discussed in the remaining part of this section.

The degree of entanglement of the initial state eq. (\ref{a1}) depends on the actual values of the parameters entering $J$. For example, in the $N=2$ case the maximal entanglement is achieved by putting $\gamma=\frac{\pi}{2}$ in eq. (\ref{a}). It is known that the resulting game possesses special properties. In fact, it has been shown that, unless some restriction on $\Sigma$ are imposed, to any move of Alice there correspond a "countermove" of Bob which allows him to neutralize Alice intentions (and vice versa) \cite{BenjaminHay}, \cite{Bolonek}. This is easily seen in the quaternionic formalism introduced by Landsburg \cite{Landsburg} \cite{Landsburg1} \cite{Landsburg2}. Since the strategies of Alice and Bob are elements of $SU(2)$ group, they can be represented by unit quaternions $q_A$ and $q_B$. It appears that the outcome probabilities eq. (\ref{b}) depend only on their product $q_A\cdot q_B$. This property makes obvious the existence of countermoves.

Our aim here is to show that the existence of countermoves is the general property of maximally entangled games even if there is no underlying  quaternionic structure (which exists only in $N=2$ case).

Let us consider a pair $\naw{U_A,U_B}$ of strategies of Alice and Bob. It is an element of $SU(N)\times SU(N)$ group. Therefore, the manifold of possible games (by the game we understand here a pair $\naw{U_A,U_B}$ of moves of Alice nad Bob) is just $SU(N)\times SU(N)$. However, one should take into account that different games may lead to the same outcome. Whether this is the case or not depends on the particular form of payoff table (for example, in the extreme case of all payoffs being equal the result of the game does not depend on the strategies chosen). There is also another, more   deep reason, related to the group geometry, for coincidence of the result   of different games. Let $G_s\in SU(N)\times SU(N)$ be the stability subgroup of the initial state $\ket{\Psi_{in}}$, i.e. the set of elements $g\in SU(N)\times SU(N)$ such that
\eq{g\ket{\Psi_{in}}=\ket{\Psi_{in}}.}
Then two games, $\naw{U_A,U_B}$ and $\naw{U_A',U_B'}$, differing by an element $g\in G_s$,
\eq{\naw{U_A',U_B'}=\naw{U_A,U_B}\cdot g}
share the same final result. The coset space $SU(N)\times SU(N)/G_s$ is the effective set of strategies.

Now the point is that $G_s$ depends on the degree of entanglement of the initial state. Consider the case of maximal entanglement. Let us write the initial state of the game as
\eq{\ket{\Psi_{in}}\equiv J\naw{\ket{1}\otimes\ket{1}}\equiv F_{ij}\ket{i}\otimes\ket{j}\label{bb},}
where the summation over repeated indices is understood and $F_{ij}=F_{ji}$.\\
The corresponding density matrix reads 
\eq{\rho_{in}=\ket{\Psi_{in}}\bra{\Psi_{in}}.}
The state described by $\rho_{in}$ is maximally entangled if the reduced density matrix is proportional to the unit matrix \cite{Bolonek2}
\eq{Tr_A\rho_{in}=\frac{1}{N}I,\qquad Tr_B\rho_{in}=\frac{1}{N}I\label{b1}.}
Eqs. (\ref{b1}) imply
\eq{FF^+=\frac{1}{N}I}
i.e. the matrix
\eq{\widetilde{F}\equiv\sqrt{N}F}
is unitary. By extracting from $\widetilde{F}$ an appropriate phase one obtains an element of $SU(N)$ group which we denote also by $\widetilde{F}$.

Let us apply an unitary transformation $U_A\otimes U_B$ to $\ket{\Psi_{in}}$:
\eq{\naw{U_A\otimes U_B}\ket{\Psi_{in}}=\naw{U_AFU_B^T}_{ij}\naw{\ket{i}\otimes\ket{j}}.}
By virtue of eq. (\ref{bb}) $\naw{U_A,U_B}\in G_s$ if
\eq{U_A\widetilde{F}U_B^T=\widetilde{F}\label{bb1}.}
The general solution to eq. (\ref{bb1}) reads
\eq{\begin{split}
& U_A=U\\
& U_B=\widetilde{F}\overline{U}\widetilde{F}^+,
\end{split}}  
where $U\in SU(N)$ is arbitrary and $\overline{U}$ denotes the complex conjugated matrix. \\
We conclude that $G_s$ consists of the elements of the form
\eq{\naw{U,\widetilde{F}\overline{U}\widetilde{F}^+}\label{f8}.}
Therefore, the stability group $G_s$ of $\ket{\Psi_{in}}$ is, up to a group automorphism, the diagonal subgroup of $SU(N)\times SU(N)$. Its Lie algebra induces the symmetric Cartan decomposition of $sU(N)\oplus sU(N)$.

Let us note that, in order to conclude that we are dealing with diagonal subgroup of $SU(N)\times SU(N)$, we don't have to assume $\widetilde {F}$ unitary. In fact, it is sufficient to take $\widetilde{F}$ invertible. Then
\begin{equation}
\naw{U,\widetilde{F}\overline{U}\widetilde{F}^{-1}}
\end{equation}
is the diagonal subgroup of $SU(N)\times SU(N)$. However, in such a case we deal with the realization of $SU(N)\times SU(N)$ with the second factor consisting of the set of matrices related by a fixed similarity transformation to the special unitary ones. They are in general no longer unitary but all relations relevant for the group-theoretical properties remain intact. However, in order to preserve the quantum-mechanical character of the game one assumes that the strategies of $\underline{\text{both}}$ players are defined by unitary matrices. Therefore, both factors of $SU(N)\times SU(N)$ must be represented by unitary matrices which calls for unitary $\widetilde{F}$ and it is this step which involves maximal entanglement assumption.

The coset manifold $SU(N)\times SU(N)/diag\naw{SU(N)\times SU(N)}$ is isomorphic as a manifold (but not a group), to the $SU(N)$ manifold. We conclude that in the case of maximal entanglement the effective set of games coincides with $SU(N)$ manifold. This allows us to write out useful decomposition of any element of $SU(N)\times SU(N)$. Let $U_1,U_2,U_A\in SU(N)$ be arbitrary; then (cf. Ref. \cite{Bolonek})
\eq{\naw{U_1,U_2}=\naw{U_A,U_2\widetilde{F}\,\overline{U}_1^+\,\overline{U}_A\widetilde{F}^+}\naw{U_A^+U_1,\widetilde{F}\,\overline{U}_A^+\,\overline{U}_1\widetilde{F}^+}\label{c}.}
The above equation can be interpreted as follows. Assume Alice choose an arbitrary strategy $U_A\in SU(N)$. Let $\naw{U_1,U_2}$ be a pair of strategies leading to the expected payoff desired by Bob. By noting that the second term on the RHS of eq. (\ref{c}) belongs to the stability group of $\ket{\Psi_{in}}$ we conclude that $U_2\widetilde{F}\,\overline{U}_1^+\,\overline{U}_A\widetilde{F}^+$ is the relevant countermove to the Alice move $U_A$.

As a result, there is no pure Nash equilibrium unless among $N^2$ pairs of classical strategies there exists one leading to the optimal outcomes both for Alice and Bob \cite{Landsburg}. In this sense there exist only trivial pure Nash equilibria.

One should stress that the existence of mixed-strategy Nash equilibria is not excluded. In fact, Nash theorem can be generalized to the quantum games \cite{Chen}. In the simplest $N=2$ case the examples of mixed-strategy Nash equilibria are given in Refs. \cite{EisertW} and \cite{FlitneyA}. 

Let us stress again that in the above reasoning neither the explicit form of payoff table nor that of gate operator $J$ are necessary; only the geometry of unitary groups enter the game.

Finally, let us note that, given a fixed classical payoff matrix, the pure Nash equilibria may not exist even if we deviate from the point of maximal entanglement. As an example consider the $N=2$ case. The relevant gate operator is given by eq. (\ref{a}) with $\gamma$ varying in the interval $\av{0,\frac{\pi}{2}}$; $\gamma =\frac{\pi}{2} $ corresponds to the maximal entanglement. Assume that apart from $t>r>p>s$ the payoffs (cf. eq. (\ref{abc})) obey $r+p>t+s$. Then no pure Nash equilibrium exists in the whole interval $\gamma_B<\gamma\leq\frac{\pi}{2}$ while for $\gamma<\gamma_B$ there is an infinite number of them; here $\sin^2\gamma_B=\frac{p-s}{\naw{p-s}+\naw{t-r}}$ \cite{DuLi3}. By taking, for example, $s=0$, $p=1$, $r=2$, $t=2+\varepsilon$ one obtains $\sin^2\gamma_B=\frac{1}{1+\varepsilon}$ so $\gamma_B$ can be arbitrary close to $\frac{\pi}{2}$. Therefore, by an appropriate choice of payoffs matrix one obtains a game possesing Nash equilibria and as close to the maximal entanglement point as one wishes. On the other hand, for any $\varepsilon>0$ the nonexistence of Nash equilibria extends to not maximally entangled games in some neighbourhood of maximally entangled one. However, the important point is that the nonexistence of Nash equilibria for maximally entangled game is of purely group-theoretical origin while otherwise the particular form of payoff matrix is relevant.

\section{Gate operators for N-startegies quantum games}

In this section we construct a wide class of entanglers (gate operators) for 2-players N-strategies quantum games. To this end one has to make some assumptions concerning the general properties of gate operator. We make only two assumptions:
\begin{description}
\item{(i)} in order to preserve the symmetry of initial classical game the gate operator $J$ is symmetric under the exchange of the factors in tensor product of Hilbert spaces ascribed to Alice and Bob;
\item{(ii)}  all classical pure strategies are contained in the set of pure quantum ones.
\end{description}  
In order (ii) to hold it is sufficient to demand the existence of $N$ matrices $U_k\in SU(N)$, $k=1,...,N$, such that (a) $U_k\ket{1}=e^{i\phi_k}\ket{k}$, $k=1,...,N$; (b) $\com{J,U_k\otimes U_l}=0$, $k,l=1,...,N$. We will further impose the conditions
\eq{\com{U_k,U_l}=0,\qquad k,l=1,...,N.}
This assumption leaves as much free room as possible for the choice of $J$ obeying (b).

In order to construct the matrices $U_k$ it is sufficient to consider the representation of the group of cyclic permutations of $12...N$. To this end consider the matrix
\begin{equation}
U=\left(\begin{array}{ccccc}
0 & 0 & \cdots\cdots & 0&1 \\
1 & 0 & \cdots\cdots & 0 &0\\
0 & 1 & \cdots\cdots & 0 &0\\
\cdots & \cdots &\cdots\cdots& \cdots&\cdots\\
0 & 0 & \cdots\cdots & 1 & 0 
\end{array}\right).
\end{equation}
Then the following properties hold:
\begin{equation}
\begin{split}
& U\ket{k}=\ket{k+1},\quad k=1,...,N-1 \\
& U\ket{N}=\ket{1}\\
& U^N=\mathbbm{1} \\
& \det U=\naw{-1}^{N-1}. 
\end{split}
\end{equation}
Let us define
\begin{equation}
U_k\equiv e^{\frac{i\pi\naw{N-1}\naw{k-1}}{N}}U^{k-1}\in SU(N),\qquad k=1,...,N.
\end{equation}
Then all $U_k$ commute and the condition (a) is obeyed with $\phi_k=\pi\frac{\naw{N-1}\naw{k-1}}{N}$.

In order to diagonalize the matrices $U_k$ it is sufficient to diagonalize $U$. The eigenvalues of $U$ are $1,\varepsilon, \varepsilon^2,...,\varepsilon^{N-1}$ with $\varepsilon=\exp\naw{\frac{2i\pi}{N}}$ being the first primitive $N$-th root of unity. It is not difficult to find the corresponding eigenvectors and the matrix $V$ diagonalizing $U$; the latter reads
\begin{equation}
V_{ik}=\frac{1}{\sqrt{N}}\overline{\varepsilon}^{\naw{i-1}\naw{k-1}},\qquad i,k=1,...,N.
\end{equation}
The necessary and sufficient condition for (b) to hold is
\begin{equation}
\com{J,I\otimes U}=0=\com{U\otimes I,J}.
\end{equation}
Let us define
\begin{equation}
\widetilde{J}\equiv\naw{V^+\otimes V^+}J\naw{V\otimes V\label{d1}}.
\end{equation}
Due to the equality $V^+UV=diag\naw{1,\varepsilon,...,\varepsilon^{N-1}}$ $\widetilde{J}$ must be diagonal. Let $\Lambda_i$, $i=1,...,N-1$ be any basis in Cartan subalgebra of $SU(N)$. Then $\widetilde{J}$ can be written as
\begin{equation}
\widetilde{J}=\exp\naw{i\sum_{k=1}^{N-1}\lambda_k\naw{\Lambda_k\otimes\Lambda_k}+i\sum_{k\neq l=1}^{N-1}\mu_{kl}\naw{\Lambda_k\otimes\Lambda_l+\Lambda_l\otimes\Lambda_k}}\label{d2}
\end{equation}
with $\lambda_k$ and $\mu_{kl}=\mu_{lk}$ real.
In defining $\widetilde{J}$ we omitted in the exponent the term $I\otimes I$ (it gives an irrelevant phase) as well as the terms $I\otimes \Lambda_k+\Lambda_k\otimes I$ (which amount to relabelling of the set of strategies).\\
Eqs. (\ref{d1}) and (\ref{d2}) provide the expression for a gate operator which depends on $N-1+{N-1 \choose 2}={N \choose 2}$ free parameters.

The above construction can be further generalized by replacing the matrix $U$ by a more general one
\begin{equation}
U=\left(\begin{array}{ccccc}
0 & 0 & \cdots\cdots & 0& e^{i\varphi_N} \\
e^{i\varphi_1} & 0 & \cdots\cdots & 0 &0\\
0 & e^{i\varphi_2} & \cdots\cdots & 0 &0\\
\cdots & \cdots &\cdots\cdots& \cdots&\cdots\\
0 & 0 & \cdots\cdots & e^{i\varphi_{N-1}} & 0 
\end{array}\right)
\end{equation}
and repeating the above reasoning with appropriate modifications. For example, the original ELW game $(N=2)$ is recovered with $\varphi_1=\pi$, $\varphi_2=0$.

We have obtained a multiparameter family of entanglers. The properties of the game depend on the actual values of parameters. In particular, one can pose the question how large is the manifold of effective games (pairs of strategies). As we saw in the previous section this manifold is isomorphic to the coset space $SU(N)\times SU(N)/G_s$, $G_s$ being the stability subgroup of $\ket{\Psi_{in}}$. Once $J$ is given, the stability group $G_s$ can be found as follows. First we determine the matrix $F$ with the help of eq. (\ref{bb}). Then we solve the invariance condition 
\eq{U_AFU_B^T=F\label{c1}.}
To this end we invoke the polar decomposition theorem which implies the following decomposition of $F$
\eq{F=UDV\label{d},}
where $U,V\in U(N)$ and $D$ hermitean, positive semidefinite and diagonal.

Eqs. (\ref{c1}) and (\ref{d}) can be combined to yield
\eq{
\begin{split}
& WDZ^+=D\\
& W\equiv U^+U_AU\in SU(N)\\
& Z\equiv V\overline{U}_BV^+\in SU(N).
\end{split}\label{ee}}
First equation (\ref{ee}) implies
\eq{\begin{split}
& WD^2W^+=D^2\\
& ZD^2Z^+=D^2.
\end{split}\label{ee1}}
Due to the fact that D is semidefinite diagonal we conclude that 
\eq{
\begin{split}
&WDW^+=D\\
& ZDZ^+=D
\end{split}\label{e2}}
and both $W$ and $Z$ have block-diagonal form corresponding to the eigenspaces of $D$. Moreover, by combining eqs. (\ref{ee}) and (\ref{e2}) one obtains
\eq{ZW^+D=D.}
Therefore, all blocks of $Z$ and $W$ corresponding to nonvanishing eigenvalues of $D$ coincide while the blocks corresponding to zero eigenvalues are independent. Having $W$ and $Z$ determined one can recover $U_A$ and $U_B$ with the help of eqs. (\ref{ee}):
\eq{\begin{split}
& U_A=UWU^+\\
& U_B=V^T\overline{Z}\,\overline{V}.
\end{split}\label{e3}}
If $F$ is invertible, $D$ is invertible as well and $W=Z$. Then eqs. (\ref{e3}) take the form
\eq{
\begin{split}
&U_A=UWU^+\\
& U_B=V^T\overline{W}\,\overline{V}.
\end{split}}
Thus the stability subgroup $G_s$ is isomorphic to  $S\naw{U(d_1)\times U(d_2)\times...\times U(d_m)}$, with $d_1,d_2,...,d_m$ being the multiplicities of the eigenvalues of $D$. \\
By noting that
\eq{dim\,S\naw{U(d_1)\times U(d_2)\times ...\times U(d_m)}=\sum_{i=1}^{m}d_i^2-1}
one finds that the dimension of the effective manifold of strategies equals
\eq{2N^2-\sum_{i=1}^md_i^2-1.}
Similar reasoning is valid if $F$ is noninvertible. However, in all cases considered below, the gate operators yield invertible $F$ matrices.

  \section{The case of three strategies}
  In the previous section a fairly general construction of entanglers for N-strategies  quantum games was described. We will now restrict our considerations to the $N=3$ case. This will allow us to give explicit characterization of the most general matrices representing classical strategies and to find explicitly the values of parameters yielding maximally entangled game. Moreover, in some cases (including those corresponding to maximal entanglement) the generators of stability subgroup $G_s$ are computed.
  Again, we start with assumptions concerning the matrices representing classical strategies. For readers convenience they are summarized in eq. (\ref{f1}) and (\ref{f2}) below
  \eq{\begin{split}
& U_k\ket{1}=e^{i\varphi_k}\ket{k},\qquad k=1,2,3\\
& \com{J,U_j\otimes U_k}=0, \qquad j,k=1,2,3
\end{split}\label{f1}}
\eq{\com{U_j,U_k}=0\label{f2}.}
The only additional assumption we make for the sake of simplicity is that  $U_1=I$. By virtue of eqs. (\ref{f1}) one finds the following general form of $U_2$ and $U_3$:
\eq{
\begin{split}
& U_2=\left (\begin{array}{ccc}
0 & \alpha & \beta\\
e^{i\varphi_2} & 0 & 0\\
0 & \overline{\beta}e^{-i\varphi_2} & -\overline{\alpha}e^{-i\varphi_2}\end{array}\right), \qquad \modu{\alpha}^2+\modu{\beta}^2=1\\
& U_3=\left (\begin{array}{ccc}
0 & \gamma & \delta\\
0 & -\overline{\delta}e^{-i\varphi_3} & \overline{\gamma}e^{-i\varphi_3}\\
e^{i\varphi_3} & 0 & 0
\end{array}\right), \qquad \modu{\gamma}^2+\modu{\delta}^2=1.\end{split}\label{f3}}
Eqs. (\ref{f2}) impose further restrictions yielding
\eq{
\begin{split}
& U_2=\left(\begin{array}{ccc}
0 & 0 & \varepsilon e^{-i\varphi_3}\\
e^{i\varphi_2} & 0 & 0\\
0 & \overline{\varepsilon}e^{i\naw{\varphi_3-\varphi_2}} & 0 \end{array}\right )\\
& U_3=\left(\begin{array}{ccc}
0 & \varepsilon e^{-i\varphi_2} & 0\\
0 & 0 & \overline{\varepsilon}e^{i\naw{\varphi_2-\varphi_3}}\\
e^{i\varphi_3} & 0 & 0 \end{array}\right),\end{split}}
where $\varepsilon$ is any cubic root from unity: in what follows we assume $\varepsilon\neq 1$.

The common eigenvectors of $U_1$, $U_2$ and $U_3$ are
\eq{\widetilde{\ket{1}}=\frac{1}{\sqrt{3}}\left(\begin{array}{c} 1 \\ e^{i\varphi_2} \\ \overline{\varepsilon}e^{i\varphi_3}\end{array}\right),\qquad \widetilde{\ket{2}}=\frac{1}{\sqrt{3}}\left(\begin{array}{c} 1 \\ \overline{\varepsilon}e^{i\varphi_2} \\ e^{i\varphi_3}\end{array}\right),\qquad \widetilde{\ket{3}}=\frac{1}{\sqrt{3}}\left(\begin{array}{c} 1 \\ \varepsilon e^{i\varphi_2} \\ \varepsilon e^{i\varphi_3}\end{array}\right).}
The corresponding eigenvalues are given in Table 2. 
\begin{table}
\caption{The eigenvalues of  $U_1$, $U_2$ and $U_3$.}
\begin{center}\begin{tabular}{|c|c|c|c|}
 \cline{1-4}
  &$ U_1$ & $U_2$ & $U_3$\\  \cline{1-4}
$\lambda_1$  & 1 & 1 & $\varepsilon$ \\ \cline{1-4}
$\lambda_2$ & 1 & $\varepsilon$ & 1 \\ \cline{1-4}
$\lambda_3$ & 1 & $\varepsilon^2$ & $\varepsilon^2$\\ 
 \hline
\end{tabular}\\
\end{center}
\end{table}
By defining
\eq{V=\frac{1}{\sqrt{3}}\left(\begin{array}{ccc}
1 & 1 & 1\\
e^{i\varphi_2} & \overline{\varepsilon} e^{i\varphi_2} & \varepsilon e^{i\varphi_2}\\
\overline{\varepsilon}e^{i\varphi_3} & e^{i\varphi_3} & \varepsilon e^{i\varphi_3}\end{array}\right),\qquad VV^+=I}
one finds
\eq{\begin{split}
& \widetilde{U}_1=V^+U_1V=I\\
& \widetilde{U}_2=V^+U_2V=diag\naw{1,\varepsilon, \varepsilon^2}\\
& \widetilde{U}_3=V^+U_3V=diag\naw{\varepsilon,1,\varepsilon^2}.
\end{split}.\label{f4}}
As in the general case considered in the previous section, the operator $\widetilde{J}$, defined by eq. (\ref{d1}), commutes with $\widetilde{U}_i$, $i=1,2,3$ and can be written in the form 
\eq{\widetilde{J}=\exp i\naw{\tau\naw{\Lambda\otimes\Lambda}+\rho\naw{\Lambda\otimes\Delta+\Delta\otimes\Lambda}+\sigma\naw{\Delta\otimes\Delta}}\label{f5},}
where $\tau$, $\rho$ and $\sigma$ are arbitrary real numbers while
\eq{
\Lambda\equiv\left(\begin{array}{ccc}
1 & 0 & 0\\
0 & -1 & 0\\
0 & 0 & 0\end{array}\right), \qquad \Delta\equiv\left(\begin{array}{ccc}
1 & 0 & 0\\
0 & 0 & 0\\
0 & 0 & -1 \end{array}\right)} 
span the Cartan subalgebra of $SU(3)$.\\
Again, writing out the general expression (\ref{f5}) we omitted the irrelevant terms $I\otimes I$, $I\otimes \Lambda +\Lambda\otimes I$ and $I\otimes \Delta +\Delta\otimes I$.

Having defined $\widetilde{J}$ one can use eq. (\ref{d1}) to compute $J$. Let us, however, note that we can work directly with the gate operator $\widetilde{J}$. In fact, by defining 
\eq{
\begin{split}
& \ket{\widetilde{k},\widetilde{l}\,}\equiv\naw{V^+\otimes V^+}\ket{k,l}\equiv\naw{V^+\otimes V^+}\naw{\ket{k}\otimes\ket{l}}\\
& \widetilde{U}_{A,B}\equiv V^+U_{A,B}V
\end{split}}
one finds that the outcome probabilities take the form:
\eq{P_{kk'}=\modu{\bra{\widetilde{k},\widetilde{k}'}\widetilde{J}^+\naw{\widetilde{U}_A\otimes\widetilde{U}_B}\widetilde{J}\ket{\widetilde{1},\widetilde{1}}}^2.}
The form of matrices $\Lambda$ and $\Delta$ has been chosen for computational simplicity. However, they can be expressed in terms of standard Gell-Mann matrices as follows
\eq{\Lambda=\lambda_3,\qquad \Delta=\frac{1}{2}\naw{\lambda_3+\sqrt{3}\lambda_8}.}
Once the gate operator is determined one looks for those values of parameters which yield the maximally entangled games. To this end we write out the reduced density matrix defined by the initial state $\ket{\Psi_{in}}$
\begin{equation}
\text{Tr}_B\rho_{in} =\frac{1}{9}\left(\begin{array}{c|c|c}
 &  e^{i\naw{3\rho+\sigma+2\tau}}+ & e^{i\naw{3\rho+2\sigma+\tau}}+\\
 3 & +e^{-i\naw{\rho +2\tau}}+ & +e^{-i\naw{2\rho+\tau}}\\
 & +e^{-i\naw{2\rho+\sigma}} & +e^{-i\naw{\rho+2\sigma}}\\ \hline
  e^{-i\naw{3\rho+\sigma+\tau}}+ &  & e^{i\naw{\sigma-\tau}}+\\
  +e^{i\naw{\rho+2\tau}}+ & 3 & +e^{-i\naw{\rho-\tau}}+\\
  +e^{2\rho+\sigma} & & +e^{i\naw{\rho-\sigma}}\\ \hline
  e^{-i\naw{3\rho+2\sigma+\tau}}+ & e^{-i\naw{\sigma-\tau}} + & \\
  + e^{i\naw{2\rho+\tau}}+ & +e^{i\naw{\rho-\tau}}+ & 3\\
  +e^{i\naw{\rho+2\sigma}} & +e^{-i\naw{\rho-\sigma}} & \\ 
  \end{array}\right) \label{f6}.
 \end{equation}
By demanding (cf. eqs. (\ref{b1}))
\eq{\text{Tr}_B\rho_{in}=\frac{1}{3}I}
we find the following sets of parameters (cf. Appendix A)
\begin{equation}\begin{split}
& \left\{ \begin{array}{c}
\tau=\rho=\sigma-\frac{2\pi}{3}\\
\sigma=\frac{2\pi}{3},\frac{8\pi}{9},\frac{10\pi}{9},\frac{4\pi}{3},\frac{14\pi}{9},\frac{16\pi}{9},2\pi \end{array}\right.\\
& \left\lbrace \begin{array}{c}
\tau=\rho=\sigma+\frac{2\pi}{3}\\
\sigma=0,\frac{2\pi}{9},\frac{4\pi}{9},\frac{2\pi}{3},\frac{8\pi}{9},\frac{10\pi}{9},\frac{4\pi}{3},\frac{14\pi}{9},\frac{16\pi}{9} \end{array}\right.\\
& \left\{ \begin{array}{c}
\tau=\sigma-\frac{2\pi}{3}\\
\rho=\sigma=\frac{2\pi}{3},\frac{8\pi}{9},\frac{10\pi}{9},\frac{4\pi}{3},\frac{14\pi}{9},\frac{16\pi}{9},2\pi\end{array}\right. \\
&\left\lbrace \begin{array}{c}
\tau=\sigma+\frac{2\pi}{3}\\
\rho=\sigma=0,\frac{2\pi}{9},\frac{4\pi}{9},\frac{2\pi}{3},\frac{8\pi}{9},\frac{10\pi}{9},\frac{4\pi}{3},\frac{14\pi}{9},\frac{16\pi}{9} \end{array}\right.\\
& \left\{ \begin{array}{c}
\rho=\sigma-\frac{2\pi}{3}\\
\tau=\sigma=\frac{2\pi}{3},\frac{8\pi}{9},\frac{10\pi}{9},\frac{4\pi}{3},\frac{14\pi}{9},\frac{16\pi}{9},2\pi \end{array}\right.\\
& \left\lbrace \begin{array}{c}
\rho=\sigma+\frac{2\pi}{3}\\
\tau=\sigma=0,\frac{2\pi}{9},\frac{4\pi}{9},\frac{2\pi}{3},\frac{8\pi}{9},\frac{10\pi}{9},\frac{4\pi}{3},\frac{14\pi}{9},\frac{16\pi}{9}. \end{array}\right. 
\end{split}\label{f7}
\end{equation}
The stability subgroup for all cases listed above is isomorphic to diagonal subgroup of $SU(3)\times SU(3)$. Eq. (\ref{f8}) implies the following form of its generators
\begin{equation}
Y\otimes I-I\otimes\widetilde{F}\overline{Y}\widetilde{F}^+\label{f9},
\end{equation}  
where $Y$ runs over all generators of $SU(3)$ (for example, Gell-Mann matrices, conventionally divided by two).

Let us remind that $\widetilde{F}$ is a symmetric matrix. Therefore
\eq{\widetilde{F}\overline{\widetilde{F}}=\widetilde{F}\widetilde{F}^+=I=\widetilde{F}^+\widetilde{F}=\overline{\widetilde{F}}\widetilde{F}\label{g}.}
Substituting
\eq{Y\rightarrow Y\mp \widetilde{F}\overline{Y}\widetilde{F}^+\equiv X}
and using eq. (\ref{g}) one easily find that the generators can be put in the form 
\eq{X\otimes I\pm I\otimes X.}
Alternatively, in order to compute the generators the direct method described in Appendix B may be used. Below we write out their explicit form for some of the solutions listed in eq. (\ref{f7}):
\begin{footnotesize}
\begin{description}
\item{(i)} $\rho=\frac{2\pi}{3}$, $\sigma=\tau=0$
\begin{equation}
\begin{split}
& G_1=\naw{\lambda_1-\sqrt{3}\lambda_2+\frac{2}{\sqrt{3}}\lambda_8}\otimes I-I\otimes \naw{\lambda_1-\sqrt{3}\lambda_2+\frac{2}{\sqrt{3}}\lambda_8}\\
& G_2=\naw{\sqrt{3}\lambda_2+\lambda_3+\lambda_4-\frac{1}{\sqrt{3}}\lambda_8}\otimes I-I\otimes \naw{\sqrt{3}\lambda_2+\lambda_3+\lambda_4-\frac{1}{\sqrt{3}}\lambda_8}\\
& G_3=\naw{\lambda_3+2\lambda_6+\frac{1}{\sqrt{3}}\lambda_8}\otimes I-I\otimes\naw{\lambda_3+2\lambda_6+\frac{1}{\sqrt{3}}\lambda_8}\\
& G_4=\naw{\lambda_2+\lambda_5}\otimes I-I\otimes \naw{\lambda_2+\lambda_5}\\
& G_5=\naw{4\lambda_2+\sqrt{3}\lambda_3+2\lambda_7-3\lambda_8}\otimes I-I\otimes\naw{4\lambda_2+\sqrt{3}\lambda_3+2\lambda_7-3\lambda_8}\\
& G_6=\naw{\lambda_1-\frac{1}{2}\lambda_4+\frac{1}{4}\lambda_6-\frac{3\sqrt{3}}{4}\lambda_7-\frac{\sqrt{3}}{2}\lambda_8}\otimes I+\\
&\qquad+I\otimes\naw{\lambda_1-\frac{1}{2}\lambda_4+\frac{1}{4}\lambda_6-\frac{3\sqrt{3}}{4}\lambda_7-\frac{\sqrt{3}}{2}\lambda_8}\\
& G_7=\naw{\lambda_2-\frac{\sqrt{3}}{2}\lambda_4-\lambda_5-\frac{\sqrt{3}}{4}\lambda_6+\frac{1}{4}\lambda_7+\frac{3}{2}\lambda_8}\otimes I+\\
&\qquad +I\otimes\naw{\lambda_2-\frac{\sqrt{3}}{2}\lambda_4-\lambda_5-\frac{\sqrt{3}}{4}\lambda_6+\frac{1}{4}\lambda_7+\frac{3}{2}\lambda_8}\\
& G_8=\naw{\lambda_3-\lambda_4-\frac{1}{2}\lambda_6-\frac{\sqrt{3}}{2}\lambda_7}\otimes I+I\otimes \naw{\lambda_3-\lambda_4-\frac{1}{2}\lambda_6-\frac{\sqrt{3}}{2}\lambda_7}
\end{split}
\end{equation}
\item{(ii)} $\sigma=\frac{2\pi}{3}$, $\rho=\tau=0$
\begin{equation}
\begin{split}
& G_1=\naw{\lambda_1-\sqrt{3}\lambda_7+\frac{2}{\sqrt{3}}\lambda_8}\otimes I-I\otimes\naw{\lambda_1-\sqrt{3}\lambda_7+\frac{2}{\sqrt{3}}\lambda_8}\\
& G_2=\naw{-\lambda_3+2\lambda_4+\frac{1}{\sqrt{3}}\lambda_8}\otimes I-I\otimes\naw{-\lambda_3+2\lambda_4+\frac{1}{\sqrt{3}}\lambda_8}\\
& G_3=\naw{-\lambda_3+\lambda_6+\sqrt{3}\lambda_7-\frac{1}{\sqrt{3}}\lambda_8}\otimes I-I\otimes\naw{-\lambda_3+\lambda_6+\sqrt{3}\lambda_7-\frac{1}{\sqrt{3}}\lambda_8}\\
& G_4=\naw{\lambda_2-\lambda_7}\otimes I-I\otimes\naw{\lambda_2-\lambda_7}\\
& G_5=\naw{\sqrt{3}\lambda_3+2\lambda_5-4\lambda_7+3\lambda_8}\otimes I-I\otimes\naw{\sqrt{3}\lambda_3+2\lambda_5-4\lambda_7+3\lambda_8}\\
& G_6=\naw{\lambda_1+\frac{1}{4}\lambda_4+\frac{3\sqrt{3}}{4}\lambda_5-\frac{1}{2}\lambda_6-\frac{\sqrt{3}}{2}\lambda_8}\otimes I+\\
& \qquad +I\otimes\naw{\lambda_1+\frac{1}{4}\lambda_4+\frac{3\sqrt{3}}{4}\lambda_5-\frac{1}{2}\lambda_6-\frac{\sqrt{3}}{2}\lambda_8}\\
& G_7=\naw{\lambda_2-\frac{\sqrt{3}}{4}\lambda_4-\frac{1}{4}\lambda_5-\frac{\sqrt{3}}{2}\lambda_6+\lambda_7+\frac{3}{2}\lambda_8}\otimes I+\\
&\qquad +I\otimes\naw{\lambda_2-\frac{\sqrt{3}}{4}\lambda_4-\frac{1}{4}\lambda_5-\frac{\sqrt{3}}{2}\lambda_6+\lambda_7+\frac{3}{2}\lambda_8}\\
&G_8=\naw{\lambda_3+\frac{1}{2}\lambda_4-\frac{\sqrt{3}}{2}\lambda_5+\lambda_6}\otimes I+I\otimes\naw{\lambda_3+\frac{1}{2}\lambda_4-\frac{\sqrt{3}}{2}\lambda_5+\lambda_6}
\end{split}
\end{equation}
\item{(iii)} $\tau=\frac{2\pi}{3}$, $\rho=\sigma=0$
\begin{equation}
\begin{split}
& G_1=\naw{\lambda_1-\frac{1}{\sqrt{3}}\lambda_8}\otimes I-I\otimes\naw{\lambda_1-\frac{1}{\sqrt{3}}\lambda_8}\\
& G_2=\naw{\lambda_5+\lambda_7}\otimes I-I\otimes\naw{\lambda_5+\lambda_7}\\
& G_3=\naw{\lambda_2+\sqrt{3}\lambda_3-2\lambda_5}\otimes I-I\otimes\naw{\lambda_2+\sqrt{3}\lambda_3-2\lambda_5}\\
& G_4=\naw{\lambda_4+\lambda_6-\frac{2}{\sqrt{3}}\lambda_8}\otimes I-I\otimes\naw{\lambda_4+\lambda_6-\frac{2}{\sqrt{3}}\lambda_8}\\
& G_5=\naw{2\lambda_3+\lambda_4-2\sqrt{3}\lambda_5-\lambda_6}\otimes I-I\otimes\naw{2\lambda_3+\lambda_4-2\sqrt{3}\lambda_5-\lambda_6}\\
& G_6=\naw{\lambda_1+\lambda_4+\lambda_6+\sqrt{3}\lambda_8}\otimes I+I\otimes\naw{\lambda_1+\lambda_4+\lambda_6+\sqrt{3}\lambda_8}\\
& G_7=\naw{\lambda_2+\frac{\sqrt{3}}{2}\lambda_4+\frac{1}{2}\lambda_5-\frac{\sqrt{3}}{2}\lambda_6-\frac{1}{2}\lambda_7}\otimes I+\\
& \qquad +I\otimes \naw{\lambda_2+\frac{\sqrt{3}}{2}\lambda_4+\frac{1}{2}\lambda_5-\frac{\sqrt{3}}{2}\lambda_6-\frac{1}{2}\lambda_7}\\
& G_8=\naw{\lambda_3+\frac{1}{2}\lambda_4+\frac{\sqrt{3}}{2}\lambda_5-\frac{1}{2}\lambda_6-\frac{\sqrt{3}}{2}\lambda_7}\otimes I+\\
&\qquad +I\otimes\naw{\lambda_3+\frac{1}{2}\lambda_4+\frac{\sqrt{3}}{2}\lambda_5-\frac{1}{2}\lambda_6-\frac{\sqrt{3}}{2}\lambda_7}
\end{split}
\end{equation}
\item{(iv)} $\rho=\frac{4\pi}{3}$, $\sigma=\tau=\frac{2\pi}{3}$
\begin{equation}
\begin{split}
& G_1=\naw{\lambda_1+\sqrt{3}\lambda_2+\frac{2}{\sqrt{3}}\lambda_8}\otimes I-I\otimes\naw{\lambda_1+\sqrt{3}\lambda_2+\frac{2}{\sqrt{3}}\lambda_8}\\
& G_2= \naw{-\sqrt{3}\lambda_2+\lambda_3+\lambda_4-\frac{1}{\sqrt{3}}\lambda_8}\otimes I-I\otimes\naw{-\sqrt{3}\lambda_2+\lambda_3+\lambda_4-\frac{1}{\sqrt{3}}\lambda_8}\\
& G_3=\naw{\lambda_3+2\lambda_6+\frac{1}{\sqrt{3}}\lambda_8}\otimes I-I\otimes\naw{\lambda_3+2\lambda_6+\frac{1}{\sqrt{3}}\lambda_8}\\
& G_4=\naw{\lambda_2+\lambda_5}\otimes I-I\otimes\naw{\lambda_2+\lambda_5}\\
& G_5=\naw{4\lambda_2-\sqrt{3}\lambda_3+2\lambda_7+3\lambda_8}\otimes I-I\otimes\naw{4\lambda_2-\sqrt{3}\lambda_3+2\lambda_7+3\lambda_8}\\
& G_6=\naw{\lambda_1-\frac{1}{2}\lambda_4+\frac{1}{4}\lambda_6+\frac{3\sqrt{3}}{4}\lambda_7-\frac{\sqrt{3}}{2}\lambda_8}\otimes I+\\
& \qquad +I\otimes\naw{\lambda_1-\frac{1}{2}\lambda_4+\frac{1}{4}\lambda_6+\frac{3\sqrt{3}}{4}\lambda_7-\frac{\sqrt{3}}{2}\lambda_8}\\
& G_7=\naw{\lambda_2+\frac{\sqrt{3}}{2}\lambda_4-\lambda_5+\frac{\sqrt{3}}{4}\lambda_6+\frac{1}{4}\lambda_7-\frac{3}{2}\lambda_8}\otimes I+\\
& \qquad +I\otimes\naw{\lambda_2+\frac{\sqrt{3}}{2}\lambda_4-\lambda_5+\frac{\sqrt{3}}{4}\lambda_6+\frac{1}{4}\lambda_7-\frac{3}{2}\lambda_8}\\
& G_8=\naw{\lambda_3-\lambda_4-\frac{1}{2}\lambda_6+\frac{\sqrt{3}}{2}\lambda_7}\otimes I+I\otimes \naw{\lambda_3-\lambda_4-\frac{1}{2}\lambda_6+\frac{\sqrt{3}}{2}\lambda_7}.
\end{split}
\end{equation}
\end{description}
\end{footnotesize}
Next, consider the case when two eigenvalues of the reduced density matrix (\ref{f6}), are equal. The necessary and sufficient conditions for this to be the case are given in Appendix B. When expessed in terms of initial parameters $\rho$, $\sigma$ and $\tau$ they become quite complicated. Therefore, we consider only the solutions with one nonvanishing parameter. They read
\begin{equation}
 \left\{\begin{array}{c}
\sigma=\tau=0\\
\rho=\frac{\pi}{3},\pi,\frac{5\pi}{3}
\end{array}\right.\qquad \left\{\begin{array}{c} \sigma=\rho=0\\ \tau=\frac{\pi}{2},\frac{3\pi}{2}\end{array}\right.\qquad \left\{\begin{array}{c} \tau=\rho=0\\ \sigma=\frac{\pi}{2},\frac{3\pi}{2}\end{array}\right.\label{h}
\end{equation} 
In all the above cases the corresponding $F$ matrix (cf. eq. (\ref{bb})) is invertible. 
Following the technique exposed in Appendix B we find, for some of the solutions listed above, the relevant generators of stability subgroups.

\begin{description}
\item{(i)} $\rho=\frac{\pi}{3}$, $\sigma=\tau=0$
\begin{equation}
\begin{split}
& G_1=\naw{\lambda_1-\sqrt{3}\lambda_2+\frac{2}{\sqrt{3}}\lambda_8}\otimes I-I\otimes \naw{\lambda_1-\sqrt{3}\lambda_2+\frac{2}{\sqrt{3}}\lambda_8}\\
&G_2=\naw{\lambda_3+\lambda_4-\sqrt{3}\lambda_5-\frac{1}{\sqrt{3}}\lambda_8}\otimes I-I\otimes \naw{\lambda_3+\lambda_4-\sqrt{3}\lambda_5-\frac{1}{\sqrt{3}}\lambda_8}\\
& G_3=\naw{\lambda_3+2\lambda_6+\frac{1}{\sqrt{3}}\lambda_8}\otimes I-I\otimes\naw{\lambda_3+2\lambda_6+\frac{1}{\sqrt{3}}\lambda_8}\\
& G_4=\naw{\sqrt{3}\lambda_1+\lambda_2-\sqrt{3}\lambda_4-\lambda_5-2\lambda_7}\otimes I+\\& \qquad +I\otimes\naw{\sqrt{3}\lambda_1+\lambda_2-\sqrt{3}\lambda_4-\lambda_5-2\lambda_7}
\end{split}
\end{equation}
\item{(ii)} $\rho=\pi$, $\sigma=\tau=0$
\begin{equation}
\begin{split}
& G_1=\naw{\lambda_1-\frac{1}{\sqrt{3}}\lambda_8}\otimes I-I\otimes\naw{\lambda_1-\frac{1}{\sqrt{3}}\lambda_8}\\
& G_2=\naw{-\lambda_3+2\lambda_4+\frac{1}{\sqrt{3}}\lambda_8}\otimes I-I\otimes\naw{-\lambda_3+2\lambda_4+\frac{1}{\sqrt{3}}\lambda_8}\\
& G_3=\naw{\lambda_3+2\lambda_6+\frac{1}{\sqrt{3}}\lambda_8}\otimes I-I\otimes\naw{\lambda_3+2\lambda_6+\frac{1}{\sqrt{3}}\lambda_8}\\
& G_4=\naw{\lambda_2-\lambda_5+\lambda_7}\otimes I+I\otimes\naw{\lambda_2-\lambda_5+\lambda_7}
\end{split}
\end{equation}
\item{(iii)} $\sigma=\frac{\pi}{2}$, $\rho=\tau=0$
\begin{equation}
\begin{split}
& G_1= \naw{\lambda_1-\lambda_2-\lambda_5-\lambda_7+\frac{1}{\sqrt{3}}\lambda_8}\otimes I-I\otimes\naw{\lambda_1-\lambda_2-\lambda_5-\lambda_7+\frac{1}{\sqrt{3}}\lambda_8}\\
& G_2=\naw{-\lambda_3+2\lambda_4+\frac{1}{\sqrt{3}}\lambda_8}\otimes I-I\otimes\naw{-\lambda_3+2\lambda_4+\frac{1}{\sqrt{3}}\lambda_8}\\
& G_3=\naw{2\lambda_2-\lambda_3+2\lambda_5+2\lambda_6+2\lambda_7-\frac{1}{\sqrt{3}}\lambda_8}\otimes I-\\
&\qquad-I\otimes \naw{2\lambda_2-\lambda_3+2\lambda_5+2\lambda_6+2\lambda_7-\frac{1}{\sqrt{3}}\lambda_8}\\
& G_4=\naw{2\lambda_1+\lambda_2+\lambda_3+\lambda_5-2\lambda_6+\lambda_7+\sqrt{3}\lambda_8}\otimes I+\\
&\qquad +I\otimes\naw{2\lambda_1+\lambda_2+\lambda_3+\lambda_5-2\lambda_6+\lambda_7+\sqrt{3}\lambda_8}
\end{split}
\end{equation}
\item{(iv)} $\tau=\frac{\pi}{2}$, $\rho=\sigma=0$
\begin{equation}
\begin{split}
& G_1=\naw{\lambda_1-\frac{1}{\sqrt{3}}\lambda_8}\otimes I-I\otimes\naw{\lambda_1-\frac{1}{\sqrt{3}}\lambda_8}\\
& G_2=\naw{\lambda_4+\lambda_6-\frac{1}{\sqrt{3}}\lambda_8}\otimes I-I\otimes\naw{\lambda_4+\lambda_6-\frac{1}{\sqrt{3}}\lambda_8}\\
& G_3=\naw{\lambda_2-\lambda_3+\lambda_5+2\lambda_6-\lambda_7-\frac{1}{\sqrt{3}}\lambda_8}\otimes I-\\
& \qquad -I\otimes\naw{\lambda_2-\lambda_3+\lambda_5+2\lambda_6-\lambda_7-\frac{1}{\sqrt{3}}\lambda_8}\\
&G_4=\naw{\lambda_2+\lambda_3+\lambda_4+\lambda_5-\lambda_6-\lambda_7}\otimes I+I\otimes \naw{\lambda_2+\lambda_3+\lambda_4+\lambda_5-\lambda_6-\lambda_7}.
\end{split}
\end{equation}
\end{description}
For generic values of $\rho$, $\sigma$ and $\tau$ which correspond to three different nonvanishing eigenvalues of the reduced density matrix (\ref{f6}) we find two commuting generators spanning the Lie algebra of $S\naw{U(1)\times U(1)\times U(1)}$.\\
As an example, consider the following values of parameters: $\rho=\frac{\pi}{2}$, $\sigma=\tau=0$. Then the relevant generators read
\begin{equation}
\begin{split}
& G_1=\naw{2\lambda_1-4\lambda_2+\lambda_3+2\lambda_4-4\lambda_5+\frac{1}{\sqrt{3}}\lambda_8}\otimes I-\\
&\qquad -I\otimes\naw{2\lambda_1-4\lambda_2+\lambda_3+2\lambda_4-4\lambda_5+\frac{1}{\sqrt{3}}\lambda_8}\\
& G_2=\naw{\lambda_3+2\lambda_6+\frac{1}{\sqrt{3}}\lambda_8}\otimes I-I\otimes\naw{\lambda_3+2\lambda_6+\frac{1}{\sqrt{3}}\lambda_8}.
\end{split}
\end{equation}

\section{Discussion}
 Let us summarize our results. We have constructed a wide class of quantum versions of 2-players N-strategies classical symmetric noncooperative games. Such a construction basically amounts to determine the entangler (gate operator) which introduces quantum correlations into the game. The only assumptions concerning the gate operator is that it preserves the symmetry of the classical game we have started with and that the classical game is faithfully represented in its quantum counterpart. The resulting gate operator depends on the number of parameters and can be expressed in terms of elements of Cartan subalgebra of $SU(N)$. Its fairly general construction, valid for any $N$, presented in Sec. III, relies of the representation of the group of cyclic permutations of $12...N$. The detailed calculations performed in Sec. IV for the $N=3$ case strongly suggest that the construction presented in Sec. III is the most general one.
 
 In the original ELW game (N=2) all classical strategies, both pure and mixed, are represented by pure quantum ones. This is no longer the case for general N. By the construction all pure classical strategies are still represented by pure quantum ones. However, as it is shown in Appendix C, the mixed classical strategies are, in general, encoded by mixed quantum ones.
 
 Some insight into the structure of the guantum game is provided by group theory. We have shown that the important role is played by the stability group $G_s$ of the initial state of the game. The effective manifold of games (pair of strategies of Alice and Bob) has been defined as the coset space $SU(N)\times SU(N)/G_s$. It should be stressed that two pairs of strategies corresponding to different points of effective manifold do not  necessarily lead to different outcomes. First, the latter may coincide due to the specific form of payoff table. Moreover, the probabilities $\modu{\av{k,k'|\Psi_{out}}}^2$ do not depend on the phase of $\ket{\Psi_{in}}$. Therefore, the definition of the stability subgroup could be generalized by including the possibility that $\ket{\Psi_{in}}$ is multiplied by an overall phase. Two games differing by an element of such generalized "stability" subgroup yield the same outcome.

Note again that in order to determine the group-theoretical structure of $G_s\subset SU(N\times SU(N))$ we don't need to work with the realization of $SU(N)\times SU(N)$ in terms of pairs of special unitary matrices, it sufficies to take two copies of SU(N) consisting of sets of matrices related by similarity transformations (in general not unitary and different for both factors of $SU(N)\times SU(N)$); the group structure remains unchanged. Only at the final step one has to invoke unitarity which again is related to the maximal entanglement assumption. 
  
 However, the definition of stability group given in the text paper is sufficient for our purposes. The most important point is that the maximally entangled game corresponds to the stability group which is basically the diagonal subgroup of $SU(N)\times SU(N)$. This allows us to show, using simple group theoretical considerations, that Bob can "neutralize" any Alice move (and vice versa). As a result, no nontrivial pure Nash equilibrium exists for maximally entangled games.
 
 For nonmaximal entanglement the relation between the degree of entanglement and the structure of stability group is rather loose. However, the following important property holds. Let us denote by $\naw{g_1,g_2}$, $g_{1,2}\in SU(N)$, the elements of stability group $G_s$ and let $Pr_1\naw{g_1,g_2}=g_1$. Then, for nonmaximal entanglement, $Pr_1G_s \varsubsetneq SU(N)$. By inspecting the reasoning presented in Sec. II we conclude that the nontrivial pure Nash equilibria are now a priori allowed and their actual existence depends on the particular choice of payoff table.
 
 General considerations presented in Sec. II and III were supported by explicit computations in $N=3$ case. The basically most general form of the gate operator was found and the values of parameters leading to maximal entanglement were determined. We gave also the explicit form of the generators of stability group $G_s$ for selected cases, including both maximal and nonmaximal entanglement.  

\begin{appendices}

\section{} 
Let us determine the values of the parameters $\tau$, $\rho$, $\sigma$ corresponding to maximal entanglement. The reduced density matrix $\text{Tr}_B\rho_{in}$ reads
\begin{equation}
\text{Tr}_B\rho_{in} =\frac{1}{9}\left(\begin{array}{c|c|c}
 &  e^{i\naw{3\rho+\sigma+2\tau}}+ & e^{i\naw{3\rho+2\sigma+\tau}}+\\
 3 & +e^{-i\naw{\rho +2\tau}}+ & +e^{-i\naw{2\rho+\tau}}\\
 & +e^{-i\naw{2\rho+\sigma}} & +e^{-i\naw{\rho+2\sigma}}\\ \hline
  e^{-i\naw{3\rho+\sigma+\tau}}+ &  & e^{i\naw{\sigma-\tau}}+\\
  +e^{i\naw{\rho+2\tau}}+ & 3 & +e^{-i\naw{\rho-\tau}}+\\
  +e^{2\rho+\sigma} & & +e^{i\naw{\rho-\sigma}}\\ \hline
  e^{-i\naw{3\rho+2\sigma+\tau}}+ & e^{-i\naw{\sigma-\tau}} + & \\
  + e^{i\naw{2\rho+\tau}}+ & +e^{i\naw{\rho-\tau}}+ & 3\\
  +e^{i\naw{\rho+2\sigma}} & +e^{-i\naw{\rho-\sigma}} & \\ 
  \end{array}\right)\label{dd2} 
 \end{equation}
 The vanishing of off-diagonal components yields
 \begin{equation}
 e^{i\naw{\alpha+\beta}}+e^{-i\alpha}+e^{-i\beta}=0\label{dd1}
 \end{equation}
 for $\alpha=\rho+2\tau$, $\beta=\sigma+2\rho$, $\alpha=2\rho+\tau$, $\beta=2\sigma+\rho$ and $\alpha=\tau-\rho$, $\beta=\rho-\sigma$. \\
 Eq. (\ref{dd1}) gives $\modu{e^{i\alpha}+e^{i\beta}}=1$ or 
 \begin{equation}
 \cos\naw{\alpha-\beta}=-\frac{1}{2}\quad i.e. \quad \alpha-\beta=\pm\frac{2\pi}{3}+2k\pi.
 \end{equation}
Inserting this back into eq. (\ref{dd1}) one arrives at six solutions (modulo $2k\pi$):
\begin{equation}\begin{split}
& (i)\quad \alpha=0,\quad \beta=\pm\frac{2\pi}{3}\\
&(ii)\quad\alpha=\pm\frac{2\pi}{3},\quad\beta=0\\
&(iii)\quad\alpha=\pm\frac{2\pi}{3}\quad\beta=\mp\frac{2\pi}{3}.
\end{split}\label{dd3}
\end{equation}
Considering the $(2,3)$-element of the matrix (\ref{dd2}) we have 
\begin{equation}
\begin{split}
& \alpha=\tau-\rho\\
& \beta=\rho-\sigma.
\end{split}
\end{equation}
Inserting here for $\alpha$ and $\beta$ the solutions (\ref{dd3}) we find $\rho$ and $\tau$ in terms of $\sigma$. This allows to determine $\sigma$ from the condition that one of the remaining off-diagonal element vanishes; it remains to check that the third element also vanishes. In this way we obtain the following solutions:
\begin{equation}\begin{split}
& \left\{ \begin{array}{c}
\tau=\rho=\sigma-\frac{2\pi}{3}\\
\sigma=\frac{2\pi}{3},\frac{8\pi}{9},\frac{10\pi}{9},\frac{4\pi}{3},\frac{14\pi}{9},\frac{16\pi}{9},2\pi \end{array}\right.\\
& \left\lbrace \begin{array}{c}
\tau=\rho=\sigma+\frac{2\pi}{3}\\
\sigma=0,\frac{2\pi}{9},\frac{4\pi}{9},\frac{2\pi}{3},\frac{8\pi}{9},\frac{10\pi}{9},\frac{4\pi}{3},\frac{14\pi}{9},\frac{16\pi}{9} \end{array}\right.\\
& \left\{ \begin{array}{c}
\tau=\sigma-\frac{2\pi}{3}\\
\rho=\sigma=\frac{2\pi}{3},\frac{8\pi}{9},\frac{10\pi}{9},\frac{4\pi}{3},\frac{14\pi}{9},\frac{16\pi}{9},2\pi\end{array}\right. \\
&\left\lbrace \begin{array}{c}
\tau=\sigma+\frac{2\pi}{3}\\
\rho=\sigma=0,\frac{2\pi}{9},\frac{4\pi}{9},\frac{2\pi}{3},\frac{8\pi}{9},\frac{10\pi}{9},\frac{4\pi}{3},\frac{14\pi}{9},\frac{16\pi}{9} \end{array}\right.\\
& \left\{ \begin{array}{c}
\rho=\sigma-\frac{2\pi}{3}\\
\tau=\sigma=\frac{2\pi}{3},\frac{8\pi}{9},\frac{10\pi}{9},\frac{4\pi}{3},\frac{14\pi}{9},\frac{16\pi}{9},2\pi \end{array}\right.\\
& \left\lbrace \begin{array}{c}
\rho=\sigma+\frac{2\pi}{3}\\
\tau=\sigma=0,\frac{2\pi}{9},\frac{4\pi}{9},\frac{2\pi}{3},\frac{8\pi}{9},\frac{10\pi}{9},\frac{4\pi}{3},\frac{14\pi}{9},\frac{16\pi}{9} \end{array}\right.
\end{split}
\end{equation}
Consider next the case of partial entanglement, i.e. the case when the matrix (\ref{dd2}) has two equal eigenvalues. In order to find the constraint on $\rho$, $\sigma$ and $\tau$ one can neglect the diagonal part of (\ref{dd2}) and consider the characteristic equation
\begin{equation}
\det\left(\begin{array}{ccc}
-\lambda & a & b\\
\overline{a} & -\lambda & c\\
\overline{b} & \overline{c} & -\lambda \end{array}\right)=0\label{dd4}
\end{equation}
where $a$, $b$ and $c$ are the off-diagonal elements of (\ref{dd2}) ($a=e^{i\naw{3\rho+\sigma+2\tau}}+e^{-i\naw{\rho+2\tau}}+e^{-i\naw{2\rho+\sigma}}$, etc.). 
Eq. (\ref{dd4}) yields
\begin{equation}
\lambda^3-\naw{\modu{a}^2+\modu{b}^2+\modu{c}^2}\lambda-\naw{a\overline{b}c+\overline{a}b\overline{c}}=0.\label{dd5}
\end{equation}
If (\ref{dd5}) has a double root
\begin{equation}
3\lambda^2-\naw{\modu{a}^2+\modu{b}^2+\modu{c}^2}=0\label{dd6}
\end{equation}
or
\begin{equation}
\lambda=\pm\sqrt{\frac{\modu{a}^2+\modu{b}^2+\modu{c}^2}{3}}.\label{dd7}
\end{equation}
Inserting this back into (\ref{dd5}) one obtains
\begin{equation}
\mp\frac{2}{3}\naw{\modu{a}^2+\modu{b}^2+\modu{c}^2}\sqrt{\frac{\modu{a}^2+\modu{b}^2+\modu{c}^2}{3}}=a\overline{b}c+\overline{a}b\overline{c}
\end{equation}
which hold for at least one choice of sign on the left hand side. Taking a square of both sides yields 
\begin{equation}
\frac{4}{27}\naw{\modu{a}^2+\modu{b}^2+\modu{c}^2}^3=4\text{Re}\naw{a\overline{b}c}^2.\label{ff9}
\end{equation}
Due to the inequality
\begin{equation}
\frac{1}{3}\naw{\modu{a}^2+\modu{b}^2+\modu{c}^2}\geq\sqrt[3]{\modu{a}^2\modu{b}^2\modu{c}^2}
\end{equation}
which is saturated if $\modu{a}=\modu{b}=\modu{c}$, one finds
\begin{equation}
\frac{1}{27}\naw{\modu{a}^2+\modu{b}^2+\modu{c}^2}^3\geq\modu{a}^2\modu{b}^2\modu{c}^2\geq\modu{a}^2\modu{b}^2\modu{c}^2\cos^2\alpha,
\end{equation}
where $\alpha=\arg a-\arg b+\arg c$. Therefore, eq. (\ref{ff9}) holds only if $\modu{a}^2=\modu{b}^2=\modu{c}^2$, $\arg a-\arg b+\arg c=0,\pi\naw{mod 2\pi}$. Then, denoting by $\lambda_0$ a duble root, one finds
\begin{equation}
\modu{a}^2=\modu{b}^2=\modu{c}^2=\lambda_0.\label{ac}
\end{equation}
The third root equals $-2\lambda_0$.

Due to the complicated structure of the elements $a$, $b$, $c$, when expressed in terms of basic parameters $\rho$, $\sigma$, $\tau$, we solve eqs. (\ref{ac}) in the special case of only one nonvanishing parameter. The resulting solutions read:
\begin{equation}
 \left\{\begin{array}{c}
\sigma=\tau=0\\
\rho=\frac{\pi}{3},\pi,\frac{5\pi}{3}
\end{array}\right.\qquad \left\{\begin{array}{c} \sigma=\rho=0\\ \tau=\frac{\pi}{2},\frac{3\pi}{2}\end{array}\right.\qquad \left\{\begin{array}{c} \tau=\rho=0\\ \sigma=\frac{\pi}{2},\frac{3\pi}{2}\end{array}\right. .
\end{equation}

\section{}
We are looking for the stability subgroup of the vector $\widetilde{J}\naw{V^+\otimes V^+}\ket{1,1}$, i.e. for all pairs of matrices $\widetilde{U}_A$, $\widetilde{U}_B$ such that
\eq{\naw{\widetilde{U}_A\otimes \widetilde{U}_B}\widetilde{J}\ket{\widetilde{1},\widetilde{1}}=\widetilde{J}\ket{\widetilde{1},\widetilde{1}}.}
The generators of $\widetilde{U}_A\otimes\widetilde{U}_B$ have the form
\eq{X\otimes I+I\otimes Y}
where $X$ and $Y$ are linear combinations of Gell-Mann matrices. Therefore, we demand
\eq{\naw{X\otimes I+I\otimes Y}\widetilde{J}\ket{\widetilde{1},\widetilde{1}}=0}
or
\eq{\widetilde{J}^{-1}\naw{X\otimes I+I\otimes Y}\widetilde{J}\ket{\widetilde{1},\widetilde{1}}=0.}
Let us denote by $\varsigma$ the tensor product transposition operator
\eq{\varsigma\naw{\ket{\psi}\otimes\ket{\phi}}=\ket{\phi}\otimes\ket{\psi}.}
Now, noting that
\eq{\varsigma\widetilde{J}\ket{\widetilde{1},\widetilde{1}}=\widetilde{J}\ket{\widetilde{1},\widetilde{1}}}
we conclude that the Lie algebra of stability subgroup is spanned by the eigenvectors of $\varsigma$, i.e. the relevant generators can be chosen in the form 
\eq{X\otimes I\pm I\otimes X.}
Therefore, it is sufficient to solve 
\eq{\widetilde{J}^{-1}\naw{X\otimes I\pm I\otimes X}\widetilde{J}\ket{\widetilde{1},\widetilde{1}}=0.}
In order to compute $\widetilde{J}^{-1}\naw{X\otimes I\pm I\otimes X}\widetilde{J}$ we consider
\eq{ Y\naw{\alpha}\equiv e^{-i\alpha\naw{A\otimes\Lambda}}\naw{X\otimes I}e^{i\alpha\naw{A\otimes\Lambda}}\label{n}}
\eq{Z\naw{\alpha}=e^{-i\alpha\naw{A\otimes\Delta}}\naw{X\otimes I}e^{i\alpha\naw{A\otimes\Delta}}}
where $A$ is an element of Cartan subalgebra of $SU(3)$. With an appropriate choice of the basis we have 
\eq{\com{A,X}=a\naw{X}X.\label{n3}}
The matrices $\Lambda$ and $\Delta$ obey
\eq{\Lambda^3-\Lambda=0,\qquad \Delta^3-\Delta=0.}
Using this and the Hausdorff formula one finds
\eq{Y\naw{\alpha}=Y_1\naw{\alpha}\otimes I+Y_2\naw{\alpha}\otimes \Lambda+Y_3\naw{\alpha}\otimes\Lambda^2.\label{n1}}
Eq. (\ref{n}) implies
\eq{\dot{Y}\naw{\alpha}=-i\naw{\com{A,Y_1}\otimes\Lambda+\com{A,Y_2}\otimes\Lambda^2+\com{A,Y_3}\otimes\Lambda}\label{n2}}
or, comparing eqs. (\ref{n1}) and (\ref{n2})
\eq{\begin{split}
& \dot{Y}_1\naw{\alpha}=0\\
& \dot{Y}_2\naw{\alpha}=-i\naw{\com{A,Y_1}+\com{A,Y_3}}\\
& \dot{Y}_3\naw{\alpha}=-i\com{A,Y_2}
\end{split}.}
So we get
\eq{\begin{split}
&Y_1\naw{\alpha}=X\\
& Y_2\naw{\alpha}=\frac{1}{2}\naw{e^{-i\alpha A}X e^{i\alpha A}-e^{i\alpha A}X e^{-i\alpha A}}\\
& Y_3\naw{\alpha}=\frac{1}{2}\naw{e^{-i\alpha A}X e^{i\alpha A}+e^{i\alpha A}X e^{-i\alpha A}-2X}
\end{split}.}
By virtue of eq. (\ref{n3}) we find finally
\eq{\begin{split}
& Y_1\naw{\alpha}=X\\
& Y_2\naw{\alpha}=-i\sin\naw{\alpha a\naw{X}}X\\
& Y_3\naw{\alpha}=\naw{\cos\naw{\alpha a\naw{X}}-1}X
\end{split}} 
and
\eq{Y\naw{\alpha}=X\otimes\naw{I-i\sin\naw{\alpha a\naw{X}}\Lambda+\naw{\cos\naw{\alpha a\naw{X}}-1}\Lambda^2}.\label{n7}}
Similarly
\eq{Z\naw{\alpha}=X\otimes\naw{I-i\sin\naw{\alpha a\naw{X}}\Delta+\naw{\cos\naw{\alpha a\naw{X}}-1}\Delta^2}.\label{n8}}
Let us put
\eq{\begin{split}
& \tau\Lambda\otimes\Lambda+\rho\naw{\Lambda\otimes\Delta+\Delta\otimes\Lambda}+\sigma\Delta\otimes\Delta=\\
& =\naw{\tau\Lambda+\rho\Delta}\otimes\Lambda+\naw{\rho\Lambda+\sigma\Delta}\otimes\Delta \equiv A_1\otimes\Lambda+A_2\otimes\Delta.\end{split}}
Therefore
\eq{\widetilde{J}=e^{iA_1\otimes\Lambda}e^{iA_2\otimes\Delta};}
using eqs (\ref{n7}) and (\ref{n8}) we find
\eq{\begin{split}
& \widetilde{J}^{-1}\naw{X\otimes I}\widetilde{J}=e^{-iA_2\otimes\Delta}e^{-iA_1\otimes\Lambda}\naw{X\otimes I}e^{iA_1\otimes\Lambda}e^{iA_2\otimes\Delta}=\\& =e^{-iA_2\otimes\Delta}\naw{X\otimes\naw{I-i\sin\naw{a_1\naw{X}}\Lambda+\naw{\cos\naw{a_1\naw{X}}-1}\Lambda^2}}e^{iA_2\otimes\Delta}=\\
& =e^{-iA_2\otimes\Delta}\naw{X\otimes I}e^{iA_2\otimes\Delta}\naw{I\otimes\naw{I-i\sin\naw{a_1\naw{X}}\Lambda+\naw{\cos\naw{a_1\naw{X}}-1}\Lambda^2}}=\\
& =X\otimes\naw{I-i\sin\naw{a_2\naw{X}}\Delta+\naw{\cos\naw{a_2\naw{X}}-1}\Delta^2}\big( I-i\sin\naw{a_1\naw{X}}\Lambda+\\
&+\naw{\cos\naw{a_1\naw{X}}-1}\Lambda^2\big)=\\
&=X\otimes\naw{\naw{I-is_2\Delta+\naw{c_2-1}\Delta^2}\naw{I-is_1\Lambda+\naw{c_1-1}\Lambda^2}}
\end{split}}
where $s_i\equiv\sin\naw{a_i\naw{X}}$, $c_i\equiv\cos\naw{a_i\naw{X}}$. Summarizing, the following relation should hold for the generators of stability subgroup
\eq{\naw{X\otimes\Omega\pm\Omega\otimes X}\naw{\ket{\widetilde{1}}\otimes\ket{\widetilde{1}}}=0}
where $\Omega$ is the matrix of the form
\eq{\Omega=\left(\begin{array}{ccc}
e^{-i\naw{a_1+a_2}} & 0 & 0\\
0 & e^{ia_1} & 0\\
0 & 0 & e^{ia_2} \end{array}\right).}

\section{}
We solve here the problem whether all classical mixed strategies can be implemented by pure quantum ones. In order to preserve the factorization property for probabilities the strategy of any player must be of the form
\begin{equation}
U=e^{i\naw{\alpha\Lambda+\beta\Delta}}
\end{equation}
The relevant probabilities of respective strategies read
\begin{equation}
p_\sigma=\modu{\bra{\widetilde{\sigma}}U\ket{\widetilde{1}}}^2
\end{equation}
or, explicitly,
\begin{equation}
\begin{split}
& p_1=\frac{1}{9}\modu{e^{i\alpha}+e^{-i\alpha+i\beta}+e^{-i\beta}}^2\\
& p_2=\frac{1}{9}\modu{e^{i\alpha}+\varepsilon^2e^{-i\alpha+i\beta}+\varepsilon e^{-i\beta}}^2\\
&  p_3=\frac{1}{9}\modu{\varepsilon^2e^{i\alpha}+e^{-i\alpha+i\beta}+\varepsilon e^{-i\beta}}^2.
\end{split}
\end{equation}
Let us call $e^{-i\naw{\alpha+\beta}}\equiv u_1$, $e^{i\naw{\beta-2\alpha}}\equiv u_2$, then
\begin{equation}
\begin{split}
& p_1=\frac{1}{9}\modu{1+u_1+u_2}^2\\
& p_2=\frac{1}{9}\modu{1+\varepsilon u_1+\varepsilon^2u_2}^2.
\end{split}\label{e}
\end{equation}
Now, $p_{1,2}$ obey $0\leq p_{1,2}\leq 1$, $0\leq p_1+p_2\leq 1$.

Let $\gamma=\frac{1}{2}\naw{\arg u_1-\arg u_2}$ (if $\gamma>\frac{\pi}{2}$ we take $\gamma\rightarrow\pi-\gamma$) and $\delta=\arg\naw{u_1+u_2}$. Then eqs. (\ref{e}) can be rewritten as
\begin{equation}
\begin{split}
& \cos^2\gamma+\cos\gamma\cos\delta=\lambda\equiv\frac{9p_1-1}{4}\\
& \cos^2\naw{\gamma+\frac{2\pi}{3}}+\cos\naw{\gamma+\frac{2\pi}{3}}\cos\delta=\mu\equiv\frac{9p_2-1}{4}
\end{split}
\end{equation}
and $-\frac{1}{4}\leq\lambda,\mu\leq 2$, $-\frac{1}{2}\leq\lambda+\mu\leq\frac{7}{4}$.
Eliminating $\cos\delta$ through
\begin{equation}
\cos\delta=\frac{\lambda-\cos^2\gamma}{\cos\gamma}\label{e1}
\end{equation}
we find cubic equation for $\text{tg}\gamma$
\begin{equation}
\naw{3-2\lambda-\mu}+2\sqrt{3}\naw{2-\lambda}\text{tg}\gamma+\naw{3-2\lambda-\mu}\text{tg}^2\gamma-2\lambda\sqrt{3}\text{tg}^3\gamma=0
\end{equation}
Solving the last equation we find $\gamma$ and then $\cos\delta$ from eq. (\ref{e1}). The solution exists if $-1\leq\cos\delta\leq 1$. One can check numerically that, in general, this is not the case. For example, taking $\lambda=-\frac{1}{8}$ and $\mu=1$ we obtain that the right hand side of eq. (\ref{e1}) is equal to $-1,12041$. 

\end{appendices}

\subsection*{Acknowledgement}
I would like to thank Professor Piotr Kosi\'nski (Department of Computer Science, Faculty of Physics and Applied Informatics, University of L\'od\'z, Poland) for helpful discussion and useful remarks. 
 This research is supported by the NCN Grant no. DEC-2012/05/D/ST2/00754.
 
 I am grateful to the anonymous referee for valuable remarks which helped to improve considerably the manuscript.


\begin{thebibliography}{99}
\bibitem{EisertWL} J. Eisert, M. Wilkens, M. Lewenstein, \emph{Phys. Rev. Lett.} \textbf{83}, 3077 (1999).
\bibitem{EisertW} J. Eisert, M. Wilkens, \emph{J. Mod. Opt.} \textbf{47}, 2543  (2000).
\bibitem{BenjaminHay} S. Benjamin, P. Hayden, \emph{Phys. Rev. Lett.} \textbf{87(6)}, 069801  (2001).
\bibitem{Meyer} D. Meyer, \emph{Phys. Rev. Lett.} \textbf{82}, 1052 (1999).
\bibitem{Marinatto} L. Marinatto, T. Weber, \emph{Phys. Lett}, \textbf{A272}, 291  (2000).
\bibitem{Benjamin} S. Benjamin, \emph{Phys. Lett.}, \textbf{A277}, 180  (2000).
\bibitem{MarinattoWeb} L. Marinatto, T. Weber, \emph{Phys. Lett.} \textbf{A 277}, 183 (2000).
\bibitem{Iqbal} A. Iqbal, A. Toor, \emph{Phys. Lett} \textbf{A280}, 249 (2001).
\bibitem{DuLi} J. Du, H. Li,X. Xu, X. Zhou, R. Han, \emph{Phys. Lett} \textbf{A289}, 9  (2001).
\bibitem{EisertWL1} J. Eisert, M. Wilkens, M. Lewenstein, \emph{Phys. Rev.Lett.} \textbf{87}, 069802  (2001).
\bibitem{FlitneyA} A. Flitney, D. Abbott, \emph{Fluct. Noise Lett.} \textbf{2}, R175  (2000).
\bibitem{Iqbal1} A. Iqbal, A. Toor, \emph{Phys. Rev.} \textbf{A65}, 052328  (2002).
\bibitem{DuLi1} J. Du, H. Li, X. Xu, M. Shi, J. Wu, X. Zhou, R. Han, \emph{Phys. Rev. Lett.} \textbf{88}, 137902 (2002).
\bibitem{Enk} S.J. van Enk, R. Pike, \emph{Phys. Rev} \textbf{A66}, 024306 (2002).
\bibitem{DuLi2} J. Du, X. Xu, H. Li, X. Zhou, R. Han, \emph{Fluct. Noise Lett.} \textbf{2}, R189 (2002).
\bibitem{Guinea} F. Guinea, M.A. Martin-Delgado, \emph{Journ. Phys.} \textbf{A36}, L197 (2003).
\bibitem{FlitneyA1} A. Flitney, D. Abbott, \emph{Proc. R. Soc. Lond.} \textbf{A459}, 2463 (2003).
\bibitem{PiotrowskiS} E. Piotrowski, J. Sladkowski, \emph{Int. Journ. Theor. Phys.} \textbf{42}, 1089 (2003).
\bibitem{DuLi3} J. Du, H. Li, X. Xu, X. Zhou, R. Han, \emph{Journ. Phys.} \textbf{A36}, 6551 (2003).
\bibitem{Zhou} L. Zhou, L. Kuang, \emph{Phys. Lett} \textbf{A315}, 426  (2003).
\bibitem{Chen} L. Chen, H. Ang, D. Kiang, L. Kwek, C. Lo, \emph{Phys. Lett.} \textbf{A316}, 317 (2003).
\bibitem{Lee} C.F. Lee, N.F. Johnson, \emph{Phys. Rev.} \textbf{A67}, 022311 (2003).
\bibitem{Shimamura} J. Shimamura, S. Ozdemir, F. Morikoshi, N. Imoto, \emph{Int. Journ. Quant. Inf.} \textbf{2}, 79 (2004).
\bibitem{Shimamura1} J. Shimamura, S. Ozdemir, F. Morikoshi, N. Imoto, \emph{Phys. Lett. A} \textbf{328}, 20 (2004).
\bibitem{Ozdemir} S. Ozdemir, J. Shimamura, N. Imoto, \emph{Phys. Lett. A} \textbf{325}, 104 (2004).
\bibitem{Ozdemir1} S. Ozdemir, J. Shimamura, F. Morikoshi, N. Imoto, \emph{Phys. Lett. A} \textbf{333}, 218  (2004).
\bibitem{Ozdemir2} S. Ozdemir, J. Shimamura, N. Imoto, \emph{New Journ. Phys.}, \textbf{9}, 43 (2007). 
\bibitem{Landsburg} S. Landsburg, \emph{Notices of the Am. Math. Soc.} \textbf{51}, 394 (2004).
\bibitem{Rosero} A. Rosero, "Classification of Quantum Symmetric Non-zero Sum $2\times 2$ Games in the Eisert Scheme", quant-phys/0402117
\bibitem{NawazT1} A. Nawaz, A. Toor, \emph{Journ. Phys.} \textbf{A37}, 11457  (2004).
\bibitem{NawazT2} A. Nawaz, A. Toor, \emph{Journ. Phys.} \textbf{A37}, 4437 (2004).
\bibitem{Iqbal2} A. Iqbal, "Studies in the theory of quantum games", quant-phys/0503176
\bibitem{FlitneyA2} A. Flitney, D. Abbott, \emph{Journ. Phys.} \textbf{A38}, 449 (2005).
\bibitem{Ichikawa} T. Ichikawa, I. Tsutsui, \emph{Ann. Phys.} \textbf{322}, 531 (2007).
\bibitem{Cheon} T. Cheon, I. Tsutsui, \emph{Phys. Lett.} \textbf{A348}, 147 (2006).
\bibitem{Patel} N. Patel, \emph{Nature} \textbf{445}, 144 (2007).
\bibitem{Ichikawa1} T. Ichikawa, I. Tsutsui, T. Cheon \emph{Journ. Phys. A: Math. and Theor.} \textbf{41}, 135303 (2008).
\bibitem{FlitneyA3} A. Flitney, L. Hollenberg, \emph{Phys. Lett.} \textbf{A363}, 381 (2007).
\bibitem{Nawaz3} A. Nawaz, "The generalized quantization schemes for games and its application to quantum information", arXiv:1012.1933
\bibitem{Landsburg1} S. Landsburg, \emph{Proc. Am. Math. Soc.} \textbf{139}, 4413 (2011).
\bibitem{Landsburg2} S. Landsburg, \emph{Wiley Encyclopedia of operations Research and Management science} (2011)
\bibitem{Schneider2} D. Schneider, \emph{Journ. Phys.} \textbf{A44}, 095301 (2011).
\bibitem{Schneider} D. Schneider, \emph{Journ. Phys.} \textbf{A45}, 085303 (2012).
\bibitem{Avishai} Y. Avishai, "Some Topics in Quantum Games", arXiv:1306.0284
\bibitem{Bolonek} K. Bolonek-Laso\'n, P. Kosi\'nski, \emph{Prog. Theor. Exp. Phys.} (7), 073A02 (2013).
\bibitem{Ramzan1} M. Ramzan, \emph{ Quant. Inf. Process.} \textbf{12}, 577 (2013).
\bibitem{Ramzan2} M. Ramzan, M. K. Khan, \emph{Fluctuation and Noise Letters} \textbf{12}, 1350025 (2013). 
\bibitem{Nawaz4} A. Nawaz, \emph{Chin. Phys. Lett.} \textbf{30(5)}, 050302 (2013).
\bibitem{Frackiewicz} P. Frackiewicz, \emph{Acta Phys. Polonica B} \textbf{44}, 29 (2013).
\bibitem{Nawaz5} A. Nawaz, "Werner-like States and Strategies form of Quantum games", arXiv:1307.5508
\bibitem{Nawaz6} A. Nawaz, \emph{J. Phys. A: Math. Theor. J. Phys.} \textbf{45}, 195304 (2012).
\bibitem{Avishay} Y. Avishai, arXiv:1402.1982 (quant.-ph.)
\bibitem{Brunner} N. Brunner, N. Linden, \emph{Nature Communications} \textbf{4}, 2057 (2013).
\bibitem{Bolonek1} K. Bolonek-Laso\'n, arXiv:1403.7731 
\bibitem{Bolonek2} K. Bolonek-Laso\'n, arXiv:1402.3932
\bibitem{Bolonek3} K. Bolonek-Laso\'n, arXiv: 1404.4454

\end{thebibliography}
\end{document}